\documentclass[11pt]{article}

\usepackage{graphicx}
\usepackage{amsmath}
\usepackage{amsfonts}
\begin{document}

\markright{Obscure Physics Quarterly}

\title{Parity non-conservation in a condensed matter system}

\author{Veit Elser\\
\\
Laboratory of Atomic and Solid State Physics\\
        Cornell University\\
        Ithaca, NY 14853-2501\\
        USA
}
\date{\today}

\maketitle

\begin{abstract}
The nuclear spin of a He$^3$ quasiparticle dissolved in superfluid He$^4$ sees an apparent magnetic field proportional to the Fermi coupling constant, the superfluid condensate density, and the electron current at the He$^3$ nucleus. Whereas the direction of the current must be parallel to the quasiparticle momentum, calculating its magnitude presents an interesting theoretical challenge because it vanishes in the Born-Oppenheimer approximation. We find the effect is too small to be observed and present our results in the hope others will be inspired to look for similar effects in other systems.
\end{abstract}

\section{Introduction}
As one of the cleanest condensed matter systems, superfluid He$^4$ is a good candidate for precision experiments. With but one exception --- the isotope He$^3$ --- nothing dissolves in superfluid He$^4$. And unlike trapped cold atom systems, superfluid He$^4$ samples are truly macroscopic and can be observed over long times. At low concentration and low temperature, a dissolved He$^3$ atom behaves as a simple quasiparticle whose only degree of freedom is its momentum with respect to the superfluid condensate. The physics of dilute He$^3$ in superfluid He$^4$ is that of an ideal, weakly interacting fermi gas.

The He$^3$ quasiparticles also have a nuclear spin. In the infinite dilution limit and at low temperatures, the nuclear spin would appear to be a completely decoupled degree of freedom, since the only mechanism of spin-``lattice" relaxation, by long wavelength phonons, is very weak. However, from a symmetry perspective, the superfluid is translationally and rotationally invariant and the quasiparticle Hamiltonian could in principle have two terms in the limit of small momentum $P$:
\begin{equation}\label{quasiHam}
H=\frac{P^2}{2 M_*}-v_0\,\mathbf{P}\cdot \boldsymbol\sigma.
\end{equation}
Here $M_*$ is the quasiparticle mass (enhanced over the nuclear mass by inertia in the superfluid flow), $\boldsymbol\sigma$ are the Pauli operators of the He$^3$ nuclear spin, and $v_0$ is a parameter. Galilean invariance rules out the second term for particles in vacuum, but this is suspended for the He$^3$ quasiparticle because the superfluid condensate defines a preferred rest frame. This term is odd under parity and $v_0$ could only be nonzero if the weak interaction played a part in its origin.

We will argue that the parity non-conserving coupling $v_0$ is indeed nonzero. Although its magnitude is far too small to be observed, even in this cleanest of condensed matter systems, it is interesting that the reach of the weak interaction extends even to the low energy properties of a condensed matter system. In particular, from Hamiltonian \eqref{quasiHam} we know that the ground state of a He$^3$ quasiparticle is a definite helicity state with nonzero momentum of magnitude $P=M_* |v_0|$.

The estimation of $v_0$ is interesting theoretically because it brings together particle, atom/molecule, and condensed matter physics. Of these the molecular physics turns out to be the most challenging because one must go beyond the Born-Oppenheimer approximation to obtain a nonzero $v_0$.

\section{Weak Hamiltonian}

The effective Hamiltonian for the coupling of a nuclear spin-${\scriptstyle \frac{1}{2}}$  $\boldsymbol\sigma$ to the atomic electrons by the weak interaction is derived in Commins and Bucksbaum \cite{commins1983weak}:
\begin{equation}\label{Hweak}
H_\mathrm{weak}=\lambda \frac{G_\mathrm{F}}{c}\left(\mathbf{j}_e\cdot \boldsymbol\sigma+i \,\mathbf{j}_e\times \boldsymbol\sigma_e\cdot \boldsymbol\sigma\right).
\end{equation}
Here $G_\mathrm{F}$ is the Fermi coupling constant, $\mathbf{j}_e$ and $\mathbf{j}_e\times \boldsymbol\sigma_e$ are respectively the electron and electron spin current densities at the nucleus, and $\lambda$ is a dimensionless constant given by the Weinberg angle $\theta_\mathrm{w}$ and the nucleon axial charge $g_A$:
\begin{equation}
\lambda=\frac{1-4\sin^2\theta_\mathrm{w}}{2\sqrt{2}}g_\mathrm{A}\approx 0.05.
\end{equation}
In our system the spin density is exactly zero and we therefore only need the electron current density term in $H_\mathrm{weak}$:
\begin{equation}
\mathbf{j}_e=\frac{1}{2m}\sum_i \left(\mathbf{p}_i\,\delta^3(\mathbf{r}_i)+\delta^3(\mathbf{r}_i)\,\mathbf{p}_i\right).
\end{equation}
Here $m$ is the electron mass, the sum is over all electrons, and the electron positions are relative to the nucleus.

\section{Naive current estimate}\label{sec:naive}

In a superfluid a nonzero fraction $f_0$ of the atoms can be treated as occupying a zero momentum condensate \cite{silver1990condensate}. In this picture the constituent pairs of electrons on the He$^4$ atoms of the condensate form a uniform electron density $2 f_0 n$, where $n$ is the density of He$^4$ atoms and $f_0\approx 9\%$  has been measured by elastic neutron scattering and Green's function Monte Carlo (summarized in \cite{silver2013momentum}). In the rest frame of a He$^3$ quasiparticle of momentum $\mathbf{P}$, where the superfluid condensate has velocity $-\mathbf{P}/M_*$, the electron current density at the He$^3$ nucleus would be
\begin{equation}\label{current}
\langle\, \mathbf{j}_e\,\rangle = -\gamma\,( 2 f_0 n)(\mathbf{P}/M_*).
\end{equation}
Here $\gamma$ is a numerical factor meant to correct important correlation effects that were left out in this analysis. To appreciate these, consider the pair of electrons on the He$^3$ atom itself. These have zero current at the nucleus and would seem to shield the nucleus from any electron current in the superfluid environment. It seems clear that $\gamma$ is less than one and probably is quite small. The primary motivation for this work was to make a convincing case that $\gamma$ is nonzero.

Taking $\gamma=1$ as a generous upper bound on the true electron current density, combining \eqref{Hweak} and \eqref{current} we arrive at the following upper bound on the parameter $v_0$ in \eqref{quasiHam}:
\begin{equation}
v_0=\left(\lambda \frac{G_\mathrm{F}}{c}\right)\left(\frac{2 f_0 n}{M_*}\right)=8\times 10^{-19}\;\mbox{m/s}.
\end{equation}
To put this number in perspective, we calculate the magnetic field $B_\mathrm{eq}$ that would produce the same NMR resonance as that produced by the weak interaction on a He$^3$ quasiparticle with a root-mean-square momentum in the superfluid corresponding to thermal equilibrium at $T=1$ K:
\begin{equation}
B_\mathrm{eq}=\frac{2 v_0\sqrt{3 M_* k_\mathrm{B} T}}{\hbar\gamma_3}=5\times 10^{-17}\,\mbox{T}.
\end{equation}
Here $\gamma_3$ is the He$^3$ gyromagnetic ratio. For the nuclear dipole magnetic fields produced by other quasiparticles to be below this value the He$^3$ concentration would have to be less than $10^{-13}$. And even if such concentrations were not prohibitive for collecting signal, it seems unlikely that a resonance frequency of order
\begin{equation}
\omega=\gamma_3 B_\mathrm{eq} = 10^{-8}\,\mbox{Hz}
\end{equation}
could ever be detected.

When the restriction of solubility is relaxed, in experiments with helium nanodroplets, a coupling of angular and linear momentum closely analogous to the $\mathbf{P}\cdot \boldsymbol\sigma$ term in \eqref{quasiHam} is realized by chiral molecules immersed in the droplets \cite{quist2002dynamics}. While the strength of the coupling is no longer dependent on the weak interaction, its detection is complicated by other factors.

\section{Pairwise scattering approximation}

Much of He$^4$ superfluid phenomenology is qualitatively reproduced when the interactions between atoms is treated in the pairwise approximation \cite{bogoliubov1947theory}. We will show that this extends to the formation of an electron current at the nucleus of an impurity He$^3$ atom.

In the pair approximation, the interaction of the He$^3$ atom with the condensate is approximated as interactions with individual He$^4$ atoms. Such an interacting pair has Hamiltonian
\begin{equation}\label{Hpair}
H_\mathrm{pair}=-\frac{\hbar^2}{2M_7}\nabla^2_{\mathbf{R}_\mathrm{cm}}-\frac{\hbar^2}{2\mu}\nabla^2_{\mathbf{R}}+H_e(\mathbf{R}),
\end{equation}
where $M_7$ and $\mu$ are the total and reduced masses of the two nuclei, $M_3$ and $M_4$, $\mathbf{R}_\mathrm{cm}$ is the center of mass of the two nuclei, $\mathbf{R}=\mathbf{R}_4-\mathbf{R}_3$ is the position of the condensate atom nucleus relative to the He$^3$ nucleus, and $H_e$ is the electron Hamiltonian for fixed, specified nuclear positions.

We will need at least two eigenstates of $H_e(\mathbf{R})$: the ground state and one (or more) excited states. Their wave functions only depend on $\mathbf{R}$ and the electron positions relative to $\mathbf{R}_3$:
\begin{equation}
\phi_0(\mathbf{r}_1,\mathbf{r}_2,\mathbf{r}_3,\mathbf{r}_4;\mathbf{R})\qquad \phi'(\mathbf{r}_1,\mathbf{r}_2,\mathbf{r}_3,\mathbf{r}_4;\mathbf{R}).
\end{equation}
The ground state $\phi_0$ is spin singlet and has zero angular momentum about $\mathbf{R}$, or of type $^1\Sigma$. We assume the same for $\phi'$ and most of our derivation of the electron current only depends on these properties. After we see how the current depends on $\phi'$ we will know which excited dimer wave functions to focus on. We do not need to keep track of the electron spins if we agree that electrons 1 and 2 are spin-up while 3 and 4 are spin-down and the wave functions are antisymmetric with respect to those electron pairs. Both wave functions are real and have the following normalization:
\begin{equation}
1=\int d^3\mathbf{r}_1\cdots d^3\mathbf{r}_4 \;\phi^2.
\end{equation}

\begin{figure}[t!]
\center{\scalebox{.6}{\includegraphics{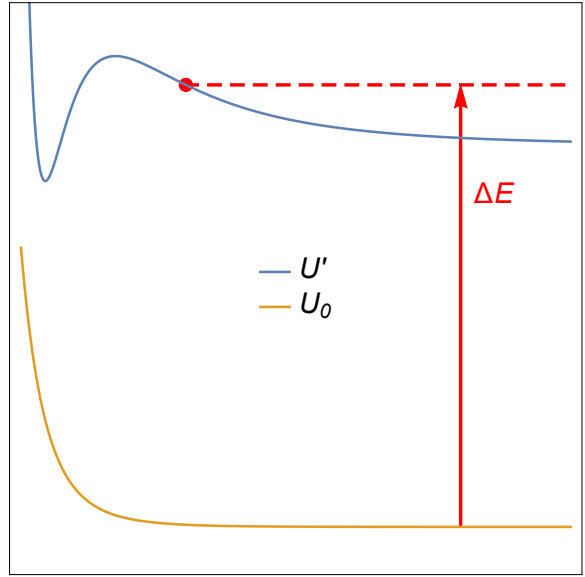}}}
\caption{Schematic rendering, based on reference \cite{guberman1975nature}, of the electronic energies $U_0(R)$ and $U'(R)$, of respectively the ground state and one excited state of the helium dimer, as a function of nuclear distance. The two curves are highly offset for clarity: for large $R$, $U'(R)-U_0(R)\sim 20\mbox{ eV}$, while the depth of the $U'(R)$ minimum is typically only about 1 eV. The excitation energy $\Delta E$, to an unbound excited-state dimer, arises in the calculation of section \ref{sec:BBO}.}
\end{figure}

Plots of the corresponding electron energies (eigenvalues of $H_e$), denoted $U_0(R)$ and $U'(R)$ as they only depend on the internuclear distance, are shown in Figure 1. The 2-body potential for nuclear motion in the electronic ground state is essentially repulsive: the very weak minimum in $U_0(R)$ (too small for the resolution in Figure 1) barely binds two He$^4$ atoms but not a He$^3$-He$^4$ pair. Whereas the potentials of the excited states usually have minima that support bound states, they share the property with $U_0(R)$ of having repulsive barriers. Only this part of the 2-body potential $U'(R)$ will play a role in the electron current.

\subsection{Born-Oppenheimer wave functions}

In the Born-Oppenheimer approximation the He$^3$-He$^4$ pair is described as two nuclei moving in the lowest energy potential, $U_0(R)$. The lowest energy wave function (of the six particle system) has zero center-of-mass momentum and zero angular momentum:
\begin{equation}
\Psi_0=\sqrt{\frac{n_0}{V}}\,\chi_0(R)\,\phi_0(\mathbf{r}_1,\mathbf{r}_2,\mathbf{r}_3,\mathbf{r}_4;\mathbf{R}).
\end{equation}
Here $\chi_0(R)$ is the nuclear wave function in the potential $U_0(R)$ with zero asymptotic nuclear kinetic energy, $n_0=f_0 n$ is the density of condensate atoms and $V$ is the system volume. The normalization factors are explained by our convention:
\begin{equation}
\chi_0(R)\sim 1,\qquad R\to \infty.
\end{equation}
From this we see that
\begin{equation}
\int_V d^3\mathbf{R}_3 \;d^3\mathbf{r}_1\cdots d^3\mathbf{r}_4 \;|\Psi_0|^2 = n_0
\end{equation}
gives the correct density of condensate atoms in the impurity atom's environment. We note that $f_0\approx 100\%$ in the dilute limit, so the detail $n_0<n$ is missed in our pairwise scattering approximation.

When the He$^3$ quasiparticle has a small momentum $\mathbf{P}$, the many-body wave function can be argued to be the following modification of the ground state wave function \cite{feynman1954atomic}:
\begin{equation}
\Psi^{(0)}_\mathbf{P}=e^{i\mathbf{P}\cdot\mathbf{R}_3/\hbar}\,\Psi_0.
\end{equation}
Being still just a multiple of a real wave function for the electrons, this Born-Oppenheimer wave function has $\langle\, \mathbf{j}_e\,\rangle=0$. We will have to apply perturbation terms from the nuclear kinetic energy that admix excited electronic wave functions $\phi'$ in order to get a nonzero current at the He$^3$ nucleus.

The excited state wave functions associated with the electronic wave function $\phi'$ have the form
\begin{equation}\label{excited}
\Psi_{\mathbf{P} k l m}=\frac{e^{i\mathbf{P}\cdot\mathbf{R}_\mathrm{cm}/\hbar}}{\sqrt{V}}\left(\sqrt{\frac{2}{R_V}} \,\chi_{k l}(R)\,Y_{lm}(\theta,\varphi)\right)\,\phi'(\mathbf{r}_1,\mathbf{r}_2,\mathbf{r}_3,\mathbf{r}_4;\mathbf{R}).
\end{equation}
Here $(R,\theta,\varphi)$ are the usual spherical coordinates of the nuclear separation $\mathbf{R}$ and $\chi_{k l}$ is a real nuclear radial wave function in the potential $U'(R)$ augmented by the centrifugal potential for angular momentum $l$. The quantum number $k$ is the  wave number associated with the asymptotic nuclear kinetic energy of the excited state with total energy $E_k$:
\begin{equation}
\frac{(\hbar k)^2}{2\mu}=E_k-U'(\infty).
\end{equation}
For large $R$,
\begin{equation}\label{largeRchi}
\chi_{k l}(R)\sim\frac{1}{R}\cos(k R-\varphi_{k l}),
\end{equation}
where the phase $\varphi_{k l}$ is determined by the vanishing of $\chi_{k l}$ for $R\to 0$. In order for $\Psi_{\mathbf{P} k l m}$ to have proper unit normalization, the radial wave functions have normalization
\begin{equation}
\frac{2}{R_V}\int_0^{R_V} \chi_{k l}^2 \;R^2 dR=1,
\end{equation}
where $R_V$ is the radius of a large bounding sphere. The wave numbers $k$ are discrete because of the boundary condition on the sphere and have density $R_V/\pi$. We see that, in the excited state wave function \eqref{excited}, the momentum $\mathbf{P}$ is shared by the scattering He$^3$-He$^4$ pair.

\subsection{Beyond Born-Oppenheimer}\label{sec:BBO}

The first correction to the  Born-Oppenheimer approximation is generated by terms in \eqref{Hpair} where the nuclear kinetic energy operator acts both on the nuclear wavefunction as well as the parametric dependence of the electronic wavefunction on the nuclear positions. Since the latter only depends on $\mathbf{R}$, we write the perturbation term as
\begin{equation}
H'=-\frac{\hbar^2}{\mu}\nabla_{\mathbf{R}_n}\cdot \nabla_{\mathbf{R}_e},
\end{equation}
where $\nabla_{\mathbf{R}_e}$ acts on $\phi_0$ and $\nabla_{\mathbf{R}_n}$ acts on the nuclear wavefunction that multiplies $\phi_0$. The corrected (six particle) wave function has the form
\begin{equation}\label{Psi(1)}
\Psi^{(1)}_\mathbf{P}=\Psi^{(0)}_\mathbf{P}-\sum_{k l m}\frac{\langle \Psi_{\mathbf{P} k l m} |H' |\Psi^{(0)}_\mathbf{P}\rangle}{E_k-E^{(0)}}\Psi_{\mathbf{P} k l m},
\end{equation}
where $E^{(0)}=U_0(\infty)$ is the energy of two separated ground-state helium atoms. We will be interested in corrections only up to linear order in the momentum $P$. At this order we will find that the sum over the angular quantum numbers only includes the term $(l=1,m=0)$, with the convention that the momentum direction $\hat{\mathbf{P}}$ defines the positive $z$-axis of our spherical coordinate system.

\subsubsection{Born-Oppenheimer matrix element}

We now evaluate the matrix element in \eqref{Psi(1)}:
\begin{align}
\mathcal{M}_{\mathbf{P} k l m}&=\langle  \Psi_{\mathbf{P} k l m} |H' |\Psi^{(0)}_\mathbf{P}\rangle\\
&=-\frac{\hbar^2}{\mu}\frac{\sqrt{n_0}}{V}\sqrt{\frac{2}{R_V}}\int d^3\mathbf{R}_3 \;d^3\mathbf{R} \;d^3\mathbf{r}_1\cdots d^3\mathbf{r}_4\label{M1}\\
& \qquad\quad \left(e^{-i\mathbf{P}\cdot\mathbf{R}_\mathrm{cm}/\hbar} \, \chi_{k l m}Y_{lm}(\theta,\varphi) \,\phi'\right)
\left(\nabla_\mathbf{R}e^{i\mathbf{P}\cdot\mathbf{R}_3/\hbar}\,\chi_0\right)\cdot\left( \nabla_\mathbf{R}\phi_0\right).\nonumber
\end{align}
Using $\mathbf{R}_3=\mathbf{R}_\mathrm{cm}-(M_4/M_7)\mathbf{R}$, the gradient acting on the nuclear wavefunction generates two terms and the integrand becomes independent of $\mathbf{R}_3$. The integrals over the electron positions,
\begin{equation}\label{u}
\int d^3\mathbf{r}_1\cdots d^3\mathbf{r}_4 \;\phi' \;\nabla_\mathbf{R}\phi_0=u(R)\,\hat{\mathbf{R}},
\end{equation}
produces a purely radial function of $\mathbf{R}$ by symmetry and defines a scalar function $u(R)$. Expanding \eqref{M1} in powers of $P$ and keeping only terms up to first order, we obtain,
\begin{align}
\mathcal{M}_{\mathbf{P} k l m}&=-\frac{\hbar^2}{\mu}\sqrt{n_0}\sqrt{\frac{2}{R_V}}\int d^3\mathbf{R} \;Y_{lm}(\theta,\varphi)\left(1-i\frac{M_4}{M_7}\left(\mathbf{P}\cdot\mathbf{R}/\hbar\right)+\cdots\right)\\
&\qquad\qquad\qquad\qquad \chi_{k l m}\left(-i\frac{M_4}{M_7}\left(\mathbf{P}\cdot\hat{\mathbf{R}}/\hbar\right)\chi_0+\frac{d \chi_0}{d R}\right) u\nonumber\\
&=i\frac{\hbar P}{M_3}\sqrt{n_0}\sqrt{\frac{2}{R_V}}\int d^3\mathbf{R} \;Y_{lm}(\theta,\varphi)\cos\theta\,\chi_{k l m} \left(\chi_0+R\,\frac{d \chi_0}{d R}\right)u,
\end{align}
where we have retained only terms proportional to $P$. For the only nonzero case, $(l=1,m=0)$, we obtain
\begin{equation}
\mathcal{M}_{\mathbf{P} k 1 0}=i\frac{\hbar P}{M_3}\sqrt{n_0}\sqrt{\frac{4\pi}{3}}\sqrt{\frac{2}{R_V}} \,A_k,
\end{equation}
where
\begin{equation}\label{A}
A_k=\int \chi_k\left(\chi_0+R\,\frac{d \chi_0}{d R}\right) u\, R^2 dR
\end{equation}
and we have dropped the $l$ and $m$ indices on $\chi_{k l m}$ since the only remaining sum is over $k$.

Recalling that $P\cos\theta=\mathbf{P}\cdot\hat{\mathbf{R}}$ in our coordinate system,  the six-particle wave function with Born-Oppenheimer correction to first order in $P$ is
\begin{align}
\Psi^{(1)}_\mathbf{P}&=\sqrt{\frac{n_0}{V}}\left(e^{i\mathbf{P}\cdot\mathbf{R}_3/\hbar}\,\chi_0 \phi_0\right.\\
&\left. \qquad -i\frac{\hbar}{M_3}\frac{2}{R_V}\sum_k \frac{A_k}{\Delta E_k}e^{i\mathbf{P}\cdot\mathbf{R}_\mathrm{cm}/\hbar}\,(\mathbf{P}\cdot\hat{\mathbf{R}})\chi_k\,\phi'\right),
\end{align}
where $\Delta E_k=E_k-U_0(\infty)$.

\subsubsection{Electron current density}

When evaluating the electron current density
\begin{equation}
\langle\,\mathbf{j}_e\,\rangle=\int d^3\mathbf{R}_3 \;d^3\mathbf{R} \;d^3\mathbf{r}_1\cdots d^3\mathbf{r}_4\;{\Psi^{(1)}_\mathbf{P}}^*\;\mathbf{j}_e\; \Psi^{(1)}_\mathbf{P}
\end{equation}
the term of order $P^0$ has a real electron wavefunction and therefore vanishing current. The cross-terms, which are of order $P^1$, also have the factors
\begin{equation}
e^{\pm i(M_4/M_7)\mathbf{P}\cdot\mathbf{R}/\hbar},
\end{equation}
which when expanded only produce higher orders in $P$. Integration of the electron positions in the cross terms produce another radially symmetric function:
\begin{equation}\label{v}
\int d^3\mathbf{r}_1\cdots d^3\mathbf{r}_4 \;\sum_{i=1}^4\delta^3(\mathbf{r}_i)\left(\phi_0\nabla_{\mathbf{r}_i} \phi'-\phi'\nabla_{\mathbf{r}_i} \phi_0\right)=v(R)\,\hat{\mathbf{R}}.
\end{equation}
This follows from the cylindrical symmetry of both $\phi_0$ and $\phi'$ about the inter-nuclear axis and because the current density is evaluated at the He$^3$ nucleus on this axis. The result of the current density calculation is
\begin{equation}\label{currentcalc}
\langle\,\mathbf{j}_e\,\rangle=-\left(\frac{4}{3}\frac{\hbar^2}{m}\int \frac{A_k B_k}{\Delta E_k}dk \right)n_0 \,(\mathbf{P}/M_3),
\end{equation}
where
\begin{equation}\label{B}
B_k=\int \chi_k\, \chi_0\, v\, R^2 dR
\end{equation}
and we used the density of wave numbers to convert the sum over $k$ into an integral.

\subsubsection{Semiclassical evaluation of nuclear integrals}

The integrals \eqref{A} and \eqref{B} involving the nuclear wave functions that define $A_k$ and $B_k$ can be simplified in the semiclassical limit, when $\chi_k$ is the only rapidly oscillating function. This is the case for the problem at hand, since the repulsive barrier in $U'(R)$ ensures the integral in \eqref{currentcalc} includes wave numbers that satisfy $k a_\mathrm{B}\gg 1$. Here the Bohr radius $a_\mathrm{B}$ represents the length scale of slow variation in the functions $u$ and $v$. The case $k\to 0$ does not present a problem because the corresponding turning point moves to large $R$ where the electronic functions $u$ and $v$ become very small. 

If maximizing the orbital mixing ($u$) or the current integral ($v$) were the only considerations, the focus would be on small $R$ and the sum over excitations would include nuclear bound states. However, the contributions to the current density from such excitations is strongly suppressed by the rapid exponential decay of $\chi_0$, for small $R$, in the repulsive potential $U_0$. The assumption of slowly varying functions in the semiclassical evaluation of nuclear integrals will also apply to $\chi_0$.

Let $R_k$ be the turning point for wave number $k$, the nuclear separation where the classical velocity is instantaneously zero when scattering with asymptotic relative momentum $\hbar k$. The semiclassical limit of integrals of $\chi_k$ with slow functions is given by just the contribution at the turning point. In the appendix we show this corresponds to the replacement
\begin{equation}\label{turningpointnorm}
\chi_k(R)\to \sqrt{\frac{\pi k}{2\mu g_k}}\;\frac{\hbar}{R_k}\;\delta(R-R_k),
\end{equation} 
where
\begin{equation}
g_k=-\left.\frac{d U'}{d R}\right|_{R_k}
\end{equation}
is the gradient at the turning point. Using this the integrals $A_k$ and $B_k$ reduce to the values of $\chi_0$, $d\chi_0/dR$, $u$ and $v$ at the turning point. Dropping the index $k$ on $R$ with the understanding that this is the turning point, we can also transform the integral over $k$ in \eqref{currentcalc} to an integral over $R$ with the Jacobian
\begin{equation}
\frac{d k}{d R}=\frac{\mu g}{\hbar^2 k}.
\end{equation}
The result of taking these steps is
\begin{equation}\label{currentfinal}
\langle\,\mathbf{j}_e\,\rangle=-\gamma\, (2n_0) (\mathbf{P}/M_3),
\end{equation}
where
\begin{equation}\label{gamma}
\gamma=\frac{\pi}{3}\frac{\hbar^2}{m}\int \left(\chi_0+R\, \frac{d\chi_0}{dR}\right)\chi_0\left(\frac{u v}{\Delta E}\right)R^2 dR
\end{equation}
no longer makes reference to the gradient $g$ at the turning point and $\Delta E=U'(R)-U_0(\infty)$ is the energy of the excited state when its nuclear wave function has turning point $R$.

\subsubsection{Excited state considerations}

To properly evaluate the dimensionless constant $\gamma$ in the current density \eqref{currentfinal}, the integral over turning points \eqref{gamma} would have to be computed for the $(u,v,\Delta E)$ of each $^1\Sigma$ excited state of the helium dimer --- their contributions add. An extensive study of the excited dimer states by Guberman and Goddard \cite{guberman1975nature} is helpful in identifying the states and integration range where we can expect the largest contributions. Since $\Delta E$ is very similar for the low lying excitations and is essentially flat (in absolute terms) as a function of $R$, the functions $u(R)$ and $v(R)$ should be our focus.

When the separation $R$ of the He$^3$-He$^4$ pair is large it is easy to see that the product $u v$ is small. In this limit, neglecting antisymmetry between electrons on different atoms, the wave functions are approximately products of 2-electron wave functions:
\begin{equation}\label{prod0}
\phi_0\approx \phi_3(\mathbf{r}_1,\mathbf{r}_2;0)\,\phi_4(\mathbf{r}_3,\mathbf{r}_4;\mathbf{R})
\end{equation}
Here $\phi_3$ and $\phi_4$ are ground state helium wave functions centered, respectively, on the He$^3$ nucleus and the He$^4$ nucleus (at $\mathbf{R}$). The Born-Oppenheimer perturbation generates the wave function
\begin{equation}
\nabla_\mathbf{R}\,\phi_0\approx \phi_3(\mathbf{r}_1,\mathbf{r}_2;0)\,\phi^*_4(\mathbf{r}_3,\mathbf{r}_4;\mathbf{R})
\end{equation}
where $\phi^*_4=\nabla_\mathbf{R}\,\phi_4$ is a combination of helium excited states with $p$-symmetry. The $p$-states on the He$^4$, when combined with the ground $s$-state (and a relative phase), generates a current --- but at the wrong nucleus. In fact, by \eqref{u} the only excited state --- again in the product approximation --- that can give a nonzero $u$ has the form
\begin{equation}\label{prodexcite}
\phi'\approx \phi_3(\mathbf{r}_1,\mathbf{r}_2;0)\,\phi'_4(\mathbf{r}_3,\mathbf{r}_4;\mathbf{R}),
\end{equation}
where $\phi'_4$ is a particular helium excitation with $p$-symmetry. But when \eqref{prodexcite} and \eqref{prod0} are used in \eqref{v} for the current at the He$^3$ nucleus, the resulting $v$ is zero.

In order for the perturbation on the He$^4$ nucleus to produce a current at the He$^3$ nucleus, the electrons on the two atoms must interact. The most direct manifestation of an interaction is the repulsive barrier in the 2-body potential $U'(R)$. Another consideration, for  current at the nucleus, is that the excited state should have $p$-type atomic character. The two lowest excited states, called A $(^1\Sigma_u)$ and C $(^1\Sigma_g)$, would appear to be ruled out by this because they correspond to a (resonating) $2s$-atomic excitation at large $R$. However, Guberman and Goddard \cite{guberman1975nature} find that the $2s$-like ``Rydberg" orbital develops $p$-like character for $R<2.2$ {\AA} in the C state, though not in the A state. The most obvious candidate is the third excited state, D $(^1\Sigma_u)$, which is $2p$-like already at large $R$. Moreover, the D state has a more sharply rising barrier, and therefore the promise of a coupling between the He$^4$ position and the He$^3$ current at larger $R$. In fact, the barrier for the D state rises in a range where $U_0$ is essentially flat and the nuclear wave function $\chi_0$ has not yet started to decay significantly.

\section{Conclusions}

The effect considered in this paper does not open a new low energy window to the weak interaction, nor promise a novel technique for measuring the elusive condensate fraction of a superfluid. As the estimate of section \ref{sec:naive} showed, the magnitude of the spin-momentum coupling is many orders of magnitude too small to be detected. The corresponding NMR frequency is so small the He$^3$ nuclear spins would only have precessed a small fraction of a period before they are randomized when the quasiparticles on which they reside scatter from the walls containing the superfluid.

What really motivated this paper was a theoretical question about the nature of superfluids. As a low energy phenomenon one automatically treats the helium atoms as single entities: the particles of the superfluid. An impurity He$^3$ is also a single entity, albeit one that is distinguishable from the other helium atoms. Lost in this abstraction is the fact that, although the two kinds of nuclei are clearly distinguishable, the electrons that surround them are not. One could describe the magnitude of the numerical factor $\gamma$ we have calculated in \eqref{gamma} as quantifying the degree to which the He$^3$ atom accepts the ``condensate electrons" of all the other helium atoms as its own. Only with respect to these shared zero-momentum electrons can the He$^3$ nucleus experience an electron wind. It is unfortunate that the weak interaction appears to be the only mechanism for detecting this wind.

\section*{Acknowledgements}

We thank Gordon Baym, Peter Lepage, Quentin Quakenbush and Cyrus Umrigar for helpful discussions.

\section{Appendix}

Here we derive the strength of the turning point contribution \eqref{turningpointnorm} to radial nuclear wave function integrals in the semiclassical limit. Let $R_k$ be the turning point of the nuclear motion in the potential $U(R)$. Near $R_k$ the nuclear wave function $\chi$ satisfies the Schr\"odinger equation
\begin{equation}\label{app1}
-\frac{\hbar^2}{2\mu}\frac{d^2\chi}{d x^2}-g_k\, x\,\chi=0,
\end{equation}
where $x=R-R_k$ and $g_k>0$ is the negative gradient of $U$ at $R_k$. The length scale
\begin{equation}
a=\left(\frac{\hbar^2}{2\mu g_k}\right)^{1/3}
\end{equation}
satisfies $a\ll a_\mathrm{B}$ for $m\ll \mu$ and lets us express $\chi$ near the turning point as the Airy function:
\begin{equation}\label{Airy}
\chi\sim (Z/a) \mathrm{Ai}(-x/a).
\end{equation}
Here $Z$ is a normalization constant to be determined and represents the strength of the turning point contribution since
\begin{equation}
\int_{-\infty}^\infty \mathrm{Ai}(y) dy = 1.
\end{equation}

The constant $Z$ can be determined by matching limits of the semiclassical approximation of $\chi$:
\begin{equation}\label{semiclass}
\chi(R)=A(R)\cos{\varphi(R)}.
\end{equation}
Here $A$ is the slowly varying amplitude and and $A^2/2$ represents the probability density. The real wave function \eqref{semiclass} corresponds to the superposition of a pair of wave packets with velocities $\pm v(R)$ where
\begin{equation}\label{velocity}
v(R)=\sqrt{\frac{2}{\mu}(U(R_k)-U(R))},
\end{equation}
and we assume $R>R_k$ in the following. The conserved radial flux of probability $j$, of one wave packet,
\begin{equation}
j=2\pi R^2 A^2 v,
\end{equation}
gives us an explicit expression for the amplitude in terms of the velocity:
\begin{equation}\label{A2}
A^2(R)=\frac{j}{2\pi R^2\, v(R)}.
\end{equation}
Since our spherical-box normalized nuclear wave function always has asymptotic form \eqref{largeRchi}, and the asymptotic velocity is $v(\infty)=\hbar k/\mu$, we infer
\begin{equation}
j=2\pi\frac{\hbar k}{\mu}.
\end{equation}
To make contact with the Airy function at the turning point, we also consider the limit of \eqref{A2} for $R-R_k=x\to 0$. Expanding \eqref{velocity} for small $x$ we then find
\begin{equation}
A(x)\sim\sqrt{\frac{\hbar k}{\mu R_k^2}}\left(\frac{\mu}{2 g_k}\right)^{1/4}\frac{1}{x^{1/4}}.
\end{equation}
Using this amplitude in \eqref{semiclass} near the turning point, comparing with \eqref{Airy} and the asymptotic behavior of the Airy function,
\begin{equation}
\mathrm{Ai}(-y)\sim\frac{1}{\sqrt{\pi}}\;\frac{1}{y^{1/4}}\cos{\varphi(y)}, \qquad y\to \infty,
\end{equation}
we obtain
\begin{equation}
Z= \sqrt{\frac{\pi k}{2\mu g_k}}\;\frac{\hbar}{R_k}.
\end{equation}

\bibliographystyle{abbrv}
\bibliography{weak}

\end{document}